\documentclass[pre,amsmath,amssymb,preprint]{revtex4}

\usepackage{amsmath}    % need for subequations
\usepackage{amsfonts}
\usepackage{amssymb}
\usepackage{graphicx}   % for figures
\usepackage{color}
\usepackage{dcolumn}
\usepackage{bm}

\begin{document}

\title{Robustness of the filamentation instability in arbitrarily oriented magnetic field: Full 3D calculation}

\author{A. Bret}
 \email{antoineclaude.bret@uclm.es}

\affiliation{ETSI Industriales, Universidad de Castilla-La Mancha, 13071 Ciudad Real, Spain}
\affiliation{Instituto de Investigaciones Energ\'{e}ticas y Aplicaciones Industriales, Campus Universitario de Ciudad Real, 13071 Ciudad Real, Spain}

\date{\today }

\begin{abstract}
The filamentation (Weibel) instability plays a key role in the formation of collisionless shocks which are thought to produce Gamma-Ray-Bursts and High-Energy-Cosmic-Rays in astrophysical environments. While it has been known for long that a flow-aligned magnetic field can completely quench the instability, it was recently proved in 2D that in the cold regime, such cancelation is possible if and only if the field is perfectly aligned. Here, this result is finally extended to a 3D geometry. Calculations are conducted for symmetric and asymmetric counter-streaming relativistic plasma shells. 2D results are retrieved in 3D: the instability can never be completely canceled for an oblique magnetic field. In addition, the maximum growth-rate is always larger for wave vectors lying in the plan defined by the flow and the oblique field. On the one hand, this bears consequences on the orientation of the generated filaments. On the other hand, it certifies 2D simulations of the problem can be performed without missing the most unstable filamentation modes.
\end{abstract}

%\pacs{52.35.Qz, 52.35.Hr, 52.50.Gj, 52.57.Kk}

\maketitle

\section{Introduction}
The filamentation instability occurring when two plasmas shells collide has attracted considerable interest in recent years \cite{califano3,HondaPRL,DeutschPRE2005,Tautz2005,BretPoPFluide,tzoufras,Rowlands2007,AllenPRL2012,Caprioli2013,Fox2013}. In Inertial Confinement Fusion (ICF), the so-called Fast Ignition Scenario \cite{Tabak} implies the unstable transport  of a relativistic electron beam passing through the fusion plasma. In astrophysics, it is thought that collisionless shocks resulting from the encounter of two collisionless plasma shells could be triggered by the same instability \cite{Medvedev1999}. The interest of these collisionless shocks is that they are able to accelerate particles up to high energies \cite{Bell1978a,Bell1978b}. Because they also host magnetic turbulence, energetic particles which have not yet escaped the shock environment radiate. These two features, acceleration and radiation in shock, could explain both High-Energy-Cosmic-Rays (HECR) and Gamma-Ray-Bursts (GRB) \cite{Piran2004}.

In collisionless conditions when the mean free path is much larger than the system size, colliding plasma shells should pass through each other without anything happening. The reason why something happens instead of nothing is that as the shells interpenetrate, the overlapping region forms a counter-streaming system notoriously unstable \cite{BretLPB2006}. When analysing the full unstable spectrum in the relativistic regime, one finds the filamentation instability tends to govern the unstable spectrum for shells of similar densities \cite{BretPRL2008}. The instability, frequently referred to as Weibel in literature, is found with a wave vector normal to the flow. Physically, it simply stems from the fact that whenever small perturbations occur that trigger the apparition of flow-aligned micro-currents in the shells, the resulting opposite currents repel each other, amplifying the perturbation.

While the occurrence of the instability is deleterious in ICF \cite{DeutschPRE2005}, its presence in astrophysical settings guarantees the formation of collisionless shocks \cite{BretPoP2013}, hence GRB's and HECR's. Great attention should then be drawn to the factors capable of quenching it.  In this respect, it has been long for now that a flow-aligned magnetic field can completely suppress the instability \cite{Godfrey1975,Cary1981,StockemPPCF2008}. But magnetic field could well be oblique with respect to the flow \cite{SironiApj}. For such configurations, it was proved recently in 2D that in the cold regime, the filamentation instability can be canceled if and only if the field is perfectly aligned with the flow \cite{BretAlvaroPoP,BretPoP2013a}. In other words, the instability cannot be suppressed by an oblique field, no matter how high. The goal of the present paper is to extend this result to the full 3D case.

\section{Basic equations}
\begin{figure}
\begin{center}
 \includegraphics[width=0.45\textwidth]{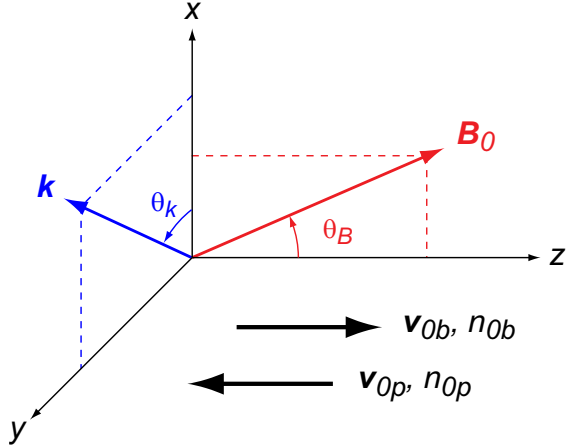}
\end{center}
\caption{(Color online) System setup and axis.}
\label{fig:1}
\end{figure}

Due to the complexity of the full kinetic theory, we address the cold (i.e. mono-kinetic) case. Our setup is sketched in Fig. \ref{fig:1}. The two homogenous and infinite shells have the initial velocities $\mathbf{v}_{0b}\parallel z$ for the rightward one, and $\mathbf{v}_{0p}\parallel z$ for the leftward. Their initial densities are $n_{0b},n_{0p}$ respectively. The magnetic field lies within the plan $(x,z)$, and the angle $\theta_B$ is such that $\mathbf{B}_0=(B_0\sin\theta_B,0,B_0\cos\theta_B)$. We study perturbations with a wave vector within the plan $(x,y)$, and the angle $\theta_k$ is defined by $\mathbf{k}=(k\cos\theta_k,k\sin\theta_k,0)$.

When the field is flow-aligned, i.e. $\theta_B=0$, the system is symmetric around the $z$ axis and one can freely align the wave vector with the $x$ axis. The growth-rate has no $\theta_k$ dependance and the system is 2D like \cite{Godfrey1975}. Here, the oblique magnetic field breaks the symmetry and the 3D problem demands accounting for the two finite perpendicular $\mathbf{k}$ components.

Both plasma shells are composed of electrons and protons, and we neglect the proton motion due to their inertia. At any rate, our goal here, besides evaluating the growth-rate, is to determine whether of not filamentation grows, and the corresponding $\theta_k$ range. The filamentation of the electrons will thus be a sign that indeed, the system is unstable and that a shock may form. Since we work in the cold regime, we can write the conservation equation and the relativistic momentum equation for both species as,
\begin{eqnarray}\label{eqs}
    \frac{\partial n_i}{\partial t}+\nabla\cdot(n_i\mathbf{v}_i)&=&0, \\
\frac{\partial \mathbf{p}_i}{\partial t}+(\mathbf{v}_i\cdot\nabla)\mathbf{p}_i&=&q\left[\mathbf{E}+\frac{\mathbf{v}_i\times(\mathbf{B}+\mathbf{B}_0)}{c}\right], \nonumber
\end{eqnarray}
with $i=p,b$ and with $\mathbf{p}_i=\gamma_i m \mathbf{v}_i$, $m$ being the electron mass.

Because both shells are charge and current neutral by their own, the whole system is charge and current neutral in any reference frame. Hence, we choose the reference frame where the initial velocities match the relation,
\begin{equation}\label{eq:ratio}
  v_{0b} = \frac{n_{0b}}{n_{0p}}v_{0p},
\end{equation}
that is, the reference frame where the center of mass is at rest. This allows to switch continuously from a symmetric collision, where plasmas come from each side at the same velocities ($n_{0b}=n_{0p}$), to the case where a diluted beam enters a plasma nearly at rest ($n_{0b}\ll n_{0p}$). In addition, working in the center of mass frame allows direct comparison with computer simulations, usually performed in this very frame.

Though lengthy, the derivation of the dispersion equation is quite standard. Equations (\ref{eqs}) are linearized assuming all quantities vary from their equilibrium value by small perturbations $\propto \exp(i\mathbf{k}\cdot \mathbf{r}-i\omega t)$. When coupled with Maxwell's equations, these linearized equations allow to derive the dielectric tensor which determinant is the dispersion equation. Calculations are conducted analytically using a \emph{Mathematica} Notebook described elsewhere  \cite{BretCPC} in terms of the variables,
\begin{equation}\label{eq:variables}
  \mathbf{Z}=\frac{\mathbf{k} v_{0b}}{\omega_{pp}},~~x = \frac{\omega}{\omega_{pp}},~~ \alpha=\frac{n_{0b}}{n_{0p}},~~\Omega_B = \frac{\omega_B}{\omega_{pp}}, ~~
  \beta=\frac{v_{0b}}{c},~~\gamma_{b0}=\frac{1}{\sqrt{1-\beta^2}},
\end{equation}
where $\omega_B=|q|B_0/mc$ is the electronic cyclotron frequency and $\omega_{pp}^2=4\pi n_{0p} q^2/m$ the electronic plasma frequency of the leftward shell. For clarity, we will treat the symmetric and diluted cases separately.

%The tensor elements are reported in Appendix for the symmetric and the diluted beam case.

\section{Validity domain}
A word of caution is needed regarding the validity of the growth-rate we are about to derive. Because the $\mathbf{B}_0$ field is not flow-aligned, it acts on the flow even at zeroth order. The present calculation are therefore meaningful provided the instability governs the early dynamic of the system, instead of the field. This implies a growth-rate $\delta$ (in $\omega_{pp}$ units) larger than the cyclotron frequency due to the transverse component of the field, i.e., 
\begin{equation}\label{eq:valid}
\delta > \Omega_B\sin\theta_B \max \left( \frac{1}{\gamma_{b0}}, \frac{1}{\gamma_{p0}} \right).
\end{equation}
where $\gamma_{b0,p0}$ refers to the Lorentz factor of the two species. For the symmetric case, both Lorentz factors are equal. In the diluted case where $n_{0b}\ll n_{0p}$, Eq. (\ref{eq:ratio}) ensures $\gamma_{p0}\sim 1$, yielding a more stringent condition (\ref{eq:valid}).

When is the inequality satisfied? In the ultra-relativistic regime  and for the symmetric case, for example, the system is like field-free. Thus, the growth-rate approaches its field-free value $\propto\gamma_{b0}^{-1/2}$ \cite{BretPoPReview}, ensuring the fulfillment of the inequality above.

Beyond the nearly trivial result, the inequality is not easily solved. We'll thus indicate on the forthcoming plots the limit it sets. Indeed, our main goal here is to find out whether or not filamentation can be quenched, and which values of $\theta_k$ are the most unstable, as long as the instability governs the early system dynamic.

\begin{figure}
\begin{center}
 \includegraphics[width=\textwidth]{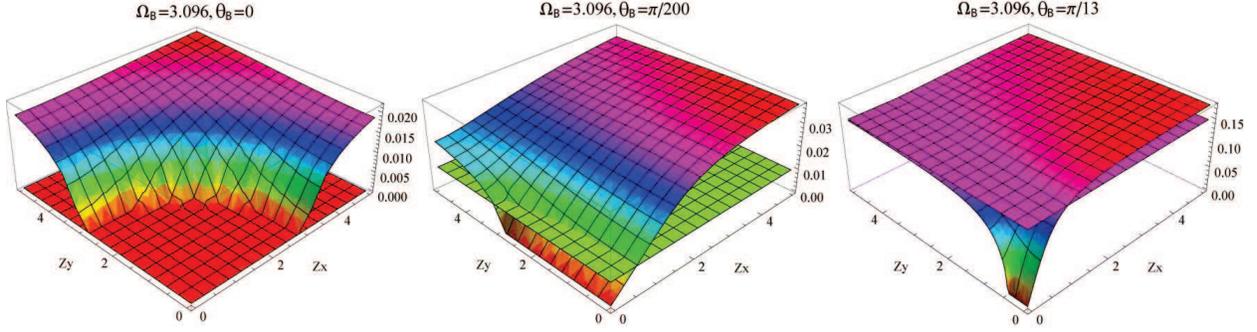}
\end{center}
\caption{(Color online) Growth-rate in $\omega_{pp}$ units, as a function of $(Z_x,Z_y)$ for 3 field obliquities and $\gamma_{b0}=5$. The horizontal plane indicates the limit sets by Eq. (\ref{eq:valid}).}
\label{fig:sym}
\end{figure}

\section{Symmetric case, $n_{0b}=n_{0p}$}
Labeling $\delta$ the growth-rate in $\omega_{pp}$ units, the symmetries $\delta(k_x,k_y)=\delta(-k_x,-k_y)$ and $\delta(k_y)=\delta(-k_y)$ allows to restrict the study to the quadrant $(k_x>0,k_y>0)$. It is known that for the 2D-like case, i.e. $\theta_k=0$, the instability vanishes for $\Omega_B>\beta\sqrt{2\gamma_{b0}}$ \cite{BretPoPMagne}. Setting $\gamma_{b0}=5$ gives a critical field amplitude $\beta\sqrt{2\gamma_{b0}}=3.098\ldots$. We thus chose $\Omega_B=3.096$ and vary the field obliquity $\theta_B$ to observe (first, numerically) how the instability cancelation evolves for $\theta_k\neq 0$. For $\theta_B=0$, one can check on figure \ref{fig:sym} how the growth-rate vanishes for $|\mathbf{Z}|\lesssim 3$. With such field orientation, the system is symmetric around the $z$ axis, and its 2D nature is recovered. Then, for $\theta_B$ as small as $\pi/200$, the growth-rate still vanishes for small $Z_y$'s but is clearly finite for other orientations.

It can be noticed that although the lower part of the plane ($Z_x,Z_y$) yields a growth-rate violating Eq. (\ref{eq:valid}), the large $Z$ part of the spectrum fulfills it and remains therefore unstable.

\begin{figure}
\begin{center}
 \includegraphics[width=0.7\textwidth]{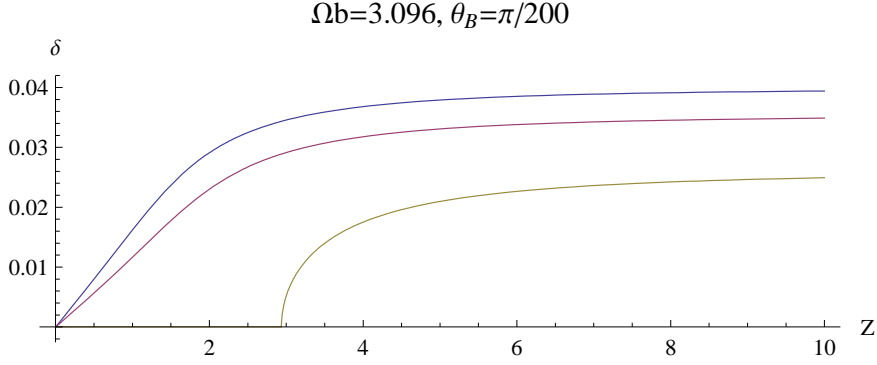}
\end{center}
\caption{(Color online) Growth-rate as a function of $|\mathbf{Z}|$ for $\theta_B=\pi/200$ and $\gamma_{b0}=5$. The angle $\theta_k$ is from bottom to top: 0, $\pi/4$ and $\pi/2$.}
\label{fig:symZ}
\end{figure}

As is the case for the 2D problem, we observe the growth-rate saturates for $Z\rightarrow\infty$. This can be checked in 3D on figure \ref{fig:symZ} where we focus on $\theta_B=\pi/200$ and plot the growth-rate in terms of $|\mathbf{Z}|$ for various orientations of the wave vector. This feature allows to write the dispersion equation in terms of $(Z_x,Z_y)=(Z\cos\theta_k,Z\sin\theta_k)$ and derive the asymptotic dispersion equation for large $Z$ from the coefficient of the higher degree in $Z$. The resulting dispersion equation at $Z=\infty$ reads,

\begin{equation}\label{eq:symZinf}
x^4 \gamma_{b0}^5
-x^2 \gamma_{b0} \left[\Omega_B^2(\sin\theta_B^2+\cos\theta_B^2 \gamma_{b0}^2) - 2 \beta ^2 \gamma_{b0}^3\right]
-\beta ^2 \Omega_B^2 \sin\theta_B^2 \cos(2\theta_k) - \beta ^2 \Omega_B^2 \sin\theta_B^2
=0.
\end{equation}
It can be solved exactly with,
\begin{eqnarray}\label{eq:delasymZinf}
\delta_{Z\infty}^2&=& \frac{ \Pi+
\sqrt{ \Pi^2 + 8\beta^2\gamma_{b0}^5\Omega_B^2\sin\theta_B^2\cos\theta_k^2}
}
{2\gamma_{b0}^5},\\
\Pi &=& 2\beta^2\gamma_{b0}^4 - \gamma_{b0}\Omega_B^2(\sin\theta_B^2+\gamma_{b0}^2\cos\theta_B^2)\nonumber.
\end{eqnarray}
Because this quantity is the high-$Z$ growth-rate, it is also the maximum growth-rate in any given direction $\theta_k$. Therefore, canceling $\delta_{Z\infty}$ implies canceling the instability. Setting $\delta_{Z\infty}=0$ now yields,
\begin{equation}
\Pi = - \sqrt{ \Pi^2 + 8\beta^2\gamma_{b0}^5\Omega_B^2\sin\theta_B^2\cos\theta_k^2},
\end{equation}
and taking the square of both side gives
\begin{equation}\label{eq:cancelsym}
 \beta\Omega_B\sin\theta_B\cos\theta_k  =  0.
\end{equation}
Note that although necessary, this condition may not be sufficient. Indeed, inserting (\ref{eq:cancelsym}) into Eq. (\ref{eq:delasymZinf}) gives,
\begin{equation}\label{eq:cancelsym1}
\delta_{Z\infty}^2= \frac{\Pi+|\Pi|}{2\gamma_{b0}^5}
\left\{ \begin{array}{ll}
=0, & \mathrm{if} ~\Pi < 0,                  \\
=\Pi/\gamma_{b0}^5, & \mathrm{if}~ \Pi > 0.
\end{array} \right.
\end{equation}
Let us now examine each possibilities offered by Eq. (\ref{eq:cancelsym}):
\begin{itemize}
  \item Eq. (\ref{eq:cancelsym}) can be fulfilled through $\beta=0$. With then $\Pi < 0$,  Eq. (\ref{eq:cancelsym1})  gives $\delta_{Z\infty}=0$. This  is the trivial case where there is no drift between the flows.
  \item Eq. (\ref{eq:cancelsym}) can be fulfilled through $\Omega_B=0$. With now $\Pi > 0$,  Eq. (\ref{eq:cancelsym1})  gives $\delta_{Z\infty}=\beta\sqrt{2/\gamma_{b0}}$. This is the maximum growth-rate for the field-free case \cite{califano3,BretPoPHierarchie}.
  \item Eq. (\ref{eq:cancelsym}) can be fulfilled through $\sin\theta_B=0$, that is, $\theta_B=0 [\pi]$. Here $\delta_{Z\infty}=0$ if $\Pi < 0$, that is, $\Omega_B>\beta\sqrt{2\gamma_{b0}}$. This is the condition for the cancellation of the instability in the flow-aligned field case \cite{BretPoPMagne}. Such cancellation is valid for all $\theta_k$, so that the entire 3D $\mathbf{k}$ spectrum is stabilized.
  \item Finally, Eq. (\ref{eq:cancelsym}) can also be fulfilled through $\cos\theta_k=0$, that is, $\theta_k=\pi/2 [\pi]$. The growth-rate vanishes now for $\Pi < 0$, namely
  \begin{equation}\label{eq.}
    \Omega_B^2 > \frac{2\beta^2\gamma_{b0}^3}{\sin\theta_B^2+\gamma_{b0}^2\cos\theta_B^2}.
  \end{equation}
  In this case, we find the instability can be canceled only for discrete orientations of the wave vector, namely, $\theta_k=\pm\pi/2$. The system remains therefore unstable for every other directions.
\end{itemize}

The proof has thus been made that it is impossible to stabilize the entire $\mathbf{k}$ spectrum unless the field is along the flow, with $\theta_B=0 [\pi]$. For an oblique field, it is quite straightforward that the growth-rate (\ref{eq:delasymZinf}) is a decreasing function of $\theta_k$ over $[0,\pi/2]$. Therefore, even when $\delta_{Z\infty}(\theta_k=\pi/2)\neq 0$, we always have $\delta_{Z\infty}(\theta_k=\pi/2) < \delta_{Z\infty}(\theta_k=0)$. This means that filamentation modes with a wave vector pertaining to the plan $(\mathbf{v}_{0b},\mathbf{B}_0)$ grow faster than the others. The orientation of the filament generated is therefore related to the orientation of the field. Also, the 2D regime (with necessarily $\theta_k=0$) usually studied in theory or simulations pertains indeed to the fastest growing configuration.

Although the large $B_0$ limit necessarily pertains to a regime governed by the field instead of the instability, the growth-rate in this limit is worth investigating in order to bridge with previous results \cite{BretAlvaroPoP,BretPoP2013a}. The expansion of Eq. (\ref{eq:delasymZinf}) is straightforward and gives,
\begin{equation}\label{eq:symZinfBinf}
\delta_{Z\infty,B_0\infty}=\beta \sqrt{\frac{2}{\gamma_{b0} }}\frac{ |\cos\theta_k|}{\sqrt{1+\gamma_{b0}^2\cot\theta_B^2}}.
\end{equation}
This result is simply the product of the field-free growth-rate for our present symmetric system, $\beta\sqrt{2/\gamma_{b0}}$, times a geometric factor. And the geometric factor is just the 2D result ($\theta_k=0$), times a $|\cos\theta_k|$ factor accounting for the orientation of $\mathbf{k}$. For $\theta_B=0$, we find $\delta_{Z\infty,B_0\infty}=0$, which means a flow-aligned field always ends-up canceling the instability. Here again, we recover the fact that the growth-rate can be canceled only for $\theta_B=0[\pi]$ or $\theta_B=\pi/2[\pi]$.

\begin{figure}
\begin{center}
 \includegraphics[width=\textwidth]{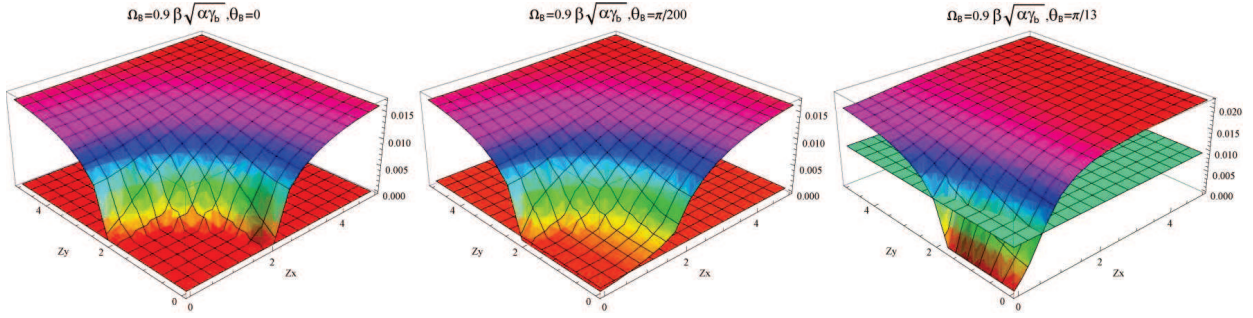}
\end{center}
\caption{(Color online) Growth-rate in $\omega_{pp}$ units, as a function of $(Z_x,Z_y)$ for 3 field obliquities, $\alpha=10^{-2}$ and $\gamma_{b0}=5$. The horizontal plane indicates the limit sets by Eq. (\ref{eq:valid}).}
\label{fig:dilu}
\end{figure}

\section{Diluted case, $n_{0b}\ll n_{0p}$}
We now treat the $n_{0b}\ll n_{0p}$, which amounts to a diluted beam passing to a plasma nearly at rest. In this respect, only the beam is relativistic now. In the diluted beam regime, the threshold for complete cancelation of the filamentation instability when $\theta_B=0$ is \cite{Godfrey1975},
\begin{equation}\label{eq:cancel_diluted}
\Omega_B>\beta\sqrt{\alpha\gamma_{b0}}.
\end{equation}
Figure \ref{fig:dilu} features the growth-rate in terms of $Z$ for various field obliquities and $\Omega_B$ close to the threshold above. The patterns observed in the symmetric case repeat here, with in particular, a saturation at large $Z$ and an uniform cancelation of the instability only achieved for $\theta_B=0$.

Still, the dispersion equation for $Z=\infty$, which gives the maximum growth-rate, in not amenable to a simple 4th order polynomial like Eq. (\ref{eq:symZinf}). Instead, we find a 8th order polynomial without simple solution. It is nevertheless possible to guess a good approximation of $\delta_{Z\infty}$ from previous works, and numerically check its relevance.

In the 2D case \cite{BretPoP2013a}, the large $B_0$ limit of the growth-rate in the diluted beam regime can be deduced from the symmetric case multiplying the result by $\sqrt{\alpha/2}$. Also, the threshold for cancelation goes from $\beta\sqrt{2\gamma_{b0}}$ in the symmetric case to $\beta\sqrt{\alpha\gamma_{b0}}$ in the diluted one. Here also, multiplying the result by $\sqrt{\alpha/2}$ allows to switch between diluted and symmetric cases. Denoting $\delta_{Z\infty}^s$ the symmetric result of Eq. (\ref{eq:delasymZinf}), the proposed \emph{ansatz} for the diluted counterpart $\delta_{Z\infty}^d$ is
\begin{equation}\label{eq:dilu_Zinf}
\delta_{Z\infty}^d = \sqrt{\frac{\alpha}{2}}   \times   \delta_{Z\infty}^s\left(\Omega_B\rightarrow\Omega_B\sqrt{\frac{2}{\alpha}}\right).
\end{equation}

\begin{figure}
\begin{center}
  \includegraphics[width=\textwidth]{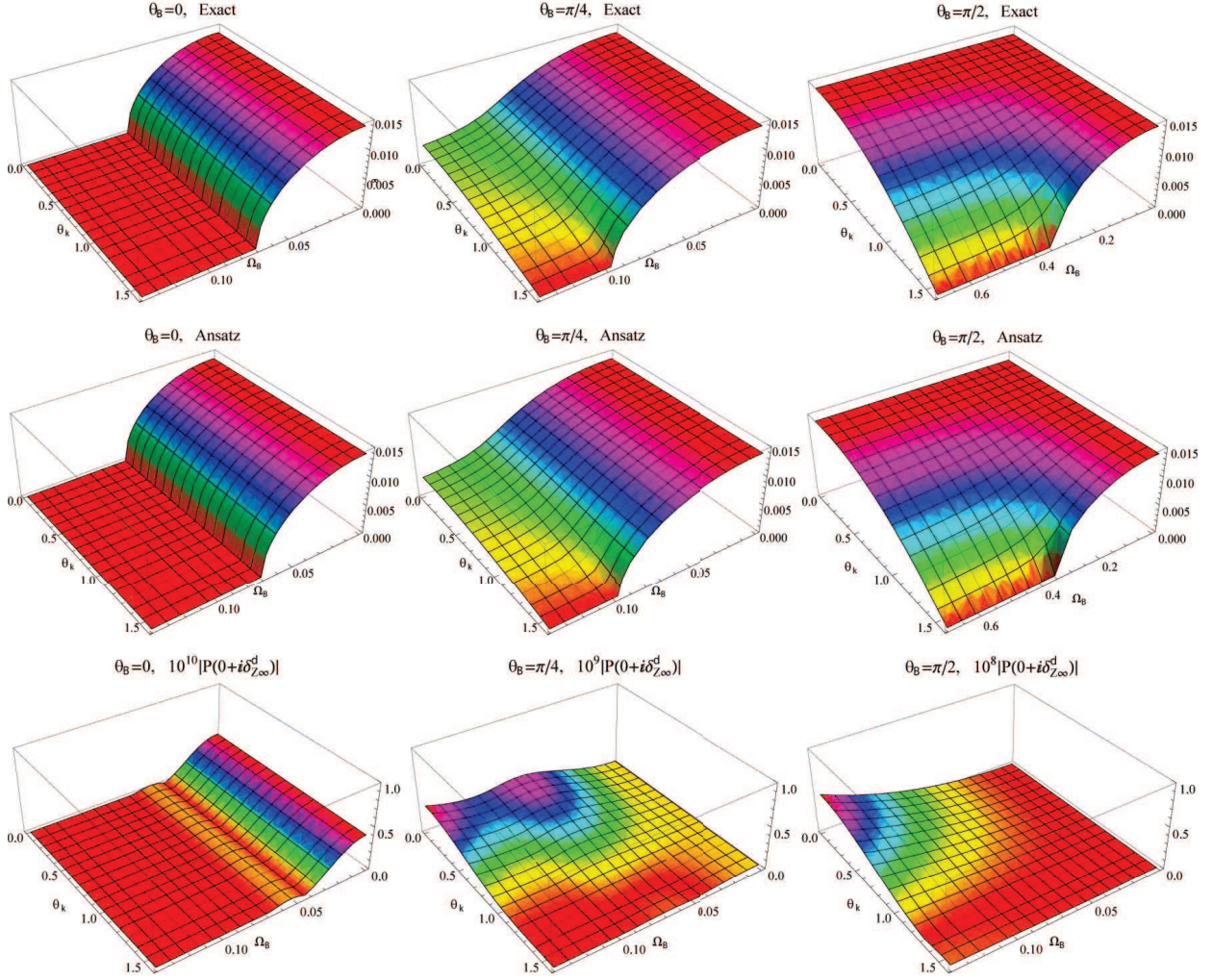}
\end{center}
\caption{(Color online) Growth-rate at $Z=\infty$ for the diluted case, exact value vs. \emph{ansatz} (\ref{eq:dilu_Zinf}). Parameters are $\alpha=10^{-3}$ and $\gamma_{b0}=5$. The last row displays the value of $\mid P(0+i\delta_{Z\infty}^d)\mid$ times a large amplification factor, where $\delta_{Z\infty}^d$ is the  \emph{ansatz} (\ref{eq:dilu_Zinf}), and $P(x)=0$ the dispersion equation.}
\label{fig:dilu_compa}
\end{figure}

Figure \ref{fig:dilu_compa} compares for various field obliquities $\theta_B$ the exact value of  $\delta_{Z\infty}^d$ with the \emph{ansatz}. The agreement is very good all over the plane $(\Omega_B,\theta_k)$.  With the dispersion equation reading $P(x)=0$, the last row displays the value of $\mid P(0+i\delta_{Z\infty}^d)\mid$, with $\delta_{Z\infty}^d$ given by Eq. (\ref{eq:dilu_Zinf}). We therefore find here a successful test of the accuracy of the \emph{ansatz}.

Here again $\theta_k=\pi/2$ is singled out as it is the only orientation for which the instability can be canceled. At any rate, the results obtained for the symmetric case are recovered. For an oblique field, the instability is canceled only for $\theta_k=\pi/2$. Furthermore, the maximum growth-rate is larger at $\theta_k=0$, which again will determine the orientation of the generated filaments.

\section{Conclusion}
The 3D filamentation instability of two cold counter-streaming beams in the presence of an oblique magnetic field has been examined for every possible orientations of the wave vector $\mathbf{k}$. Our main result is that as is the case for previous 2D treatments \cite{BretAlvaroPoP,BretPoP2013a}, the instability can never be completely quenched for a non flow-aligned field. Although the formalism is only valid when filamentation does govern the initial beam dynamic, the result remains. Results are rigorous for the symmetric beams case, and rely on an \emph{ansatz} for the diluted one.

This 3D study brings an element by definition out of reach of the 2D problem: we find a systematically larger growth-rate for $\theta_k=0$. This bears consequences on the orientation of the filaments with respect to the field and the flow. Also, 2D simulations of the system with $\mathbf{k}\in (\mathbf{v}_{0b}, \mathbf{B}_0)$ can be performed without missing the most unstable mode.

The regime where the early dynamic imposed by the field can compete with the instability is out of reach of the present study and could definitely be the subject of forthcoming works. Also, temperature effects should be investigated. While the field tends to stabilize first the long wavelengths \cite{Godfrey1975}, temperature stabilizes the short ones. It could be that with the help of kinetic effects, and oblique field can still cancel filamentation within some parameters range yet to determine.

The case where one of the two shells is composed of electrons/positrons instead of electrons/protons is worth investigating, in particular in connection with the problem of blazar-induced pair beams propagating in the intergalactic medium \cite{Schlickeiser2012a,Schlickeiser2013,Miniati2013,Sironi2013}. At present time, it seems difficult to qualitatively predict the behavior of such a system  because electrons and positrons react differently to the magnetic field. Future works will have to clarify this issue.

%\bibliography{BibBret}

\end{document}